\documentclass[12pt]{amsart}

\usepackage{amsfonts,amsmath,amsthm,amssymb}
\usepackage{latexsym}
\usepackage{graphicx}

\newtheorem{theorem}{Theorem}[section]

\newtheorem{proposition}[theorem]{Proposition}
\newtheorem{definition}{Definition}[section]
\theoremstyle{definition}

\newcommand{\Z}{\ensuremath{\mathbb{Z}}}

\def \C {\Gamma}

\newcommand \vt {{\bf t}}
\newcommand \vs {{\bf s}}
\newcommand \vg {{\bf g}}
\newcommand \vh {{\bf h}}
\newcommand \vl {{\bf l}}
\newcommand \vq {{\bf q}}

\newcommand \vr {{\bf r}}
\newcommand \vc {{\bf c}}
\newcommand \va {{\bf a}}

\def \< {\langle}
\def \> {\rangle}

\newcommand{\beql}[1]{\begin{equation}\label{#1}}
\newcommand{\eeq}{\end{equation}}
\newcommand{\comment}[1]{}

\newcommand{\RR}{{\mathbb R}}

\newcommand{\ZZ}{{\mathbb Z}}

\newcommand{\TT}{{\mathbb T}}

\newcommand{\ft}[1]{\widehat{#1}}

\newcounter{rem}
\begin{document}

\title{Constructions of complex Hadamard matrices via tiling Abelian groups}

\author{M\'at\'e Matolcsi \& J\'ulia R\'effy \& Ferenc Sz\"oll\H osi}

\date{May, 2006}

\address{M\'at\'e Matolcsi: Alfr\'ed R\'enyi Institute of Mathematics,
Hungarian Academy of Sciences POB 127 H-1364 Budapest,
Hungary.}\email{matomate@renyi.hu}

\address{J\'ulia R\'effy: Technical University Budapest (BME), Institute of
Mathematics, Department of Analysis}\email{reffyj@math.bme.hu}

\address{Ferenc Sz\"oll\H osi: Technical University Budapest (BME)}\email{szoferi@math.bme.hu}
\thanks{M. Matolcsi was supported by OTKA-T047276, T049301, PF64061. J. R\'effy was supported by OTKA-TS049835, T0466599.}

\begin{abstract}
Applications in quantum information theory and quantum tomography
have raised current interest in {\it complex Hadamard matrices}.
In this note we investigate the connection between tiling of Abelian
groups and constructions of complex Hadamard matrices. First, we
recover a recent very general construction of complex Hadamard
matrices due to Dita \cite{dita} via a natural tiling
construction. Then we find some necessary conditions for any given
complex Hadamard matrix to be {\it equivalent} to a Dita-type
matrix. Finally, using another tiling construction, due to Szab\'o
\cite{szabo}, we arrive at {\it new parametric families} of
complex Hadamard matrices of order 8, 12 and 16, and we use our
necessary conditions to prove that these families do not arise
with Dita's construction. These new families {\it complement the
recent catalogue} \cite{karol} of complex Hadamard matrices of
small order.

\end{abstract}

\maketitle

{\bf 2000 Mathematics Subject Classification.} Primary 05B20, secondary 52C22.

{\bf Keywords and phrases.} {\it Complex Hadamard matrices, spectral sets, tiling Abelian groups.}

\section{Introduction}\label{sec:intro}

Hadamard matrices, real or complex, appear in various branches of
mathematics such as combinatorics, Fourier analysis and quantum
information theory. Various applications in quantum information
theory have raised recent interest in {\it complex} Hadamard
matrices.

One example, taken from quantum tomography, is the problem of
existence of {\it mutually unbiased bases}, which is known to be a
question on the existence of certain complex Hadamard matrices.
The existence of $d+1$ such bases is known for any prime power
dimension $d$, but the problem remains open for all non prime
power dimensions, even for $d=6$ (for a more detailed exposition
of this example see the Introduction of \cite{karol}).

Other important questions in quantum information theory, such as
construction of teleportation and dense coding schemes, are also
based on complex Hadamard matrices. Werner in \cite{werner} proved
that the construction of {\it bases of maximally entangled
states}, {\it orthonormal bases of unitary operators}, and {\it
unitary depolarizers} are all equivalent in the sense that a
solution to any of them leads to a solution to any other, as well
as to a corresponding scheme of teleportation and dense coding. A
general construction procedure for orthonormal bases of unitaries,
involving complex Hadamard matrices, is also presented in
\cite{werner}.

On the one hand, it seems to be impossible to give any complete,
or satisfactory {\it characterization} of complex Hadamard
matrices of high order. On the other hand, we can hope to give
fairly {\it general constructions} producing large families of
Hadamard matrices, and we can also hope to characterize Hadamard
matrices of small order (currently a full characterization is
available only up to order 5; very recently the self-adjoint complex Hadamard matrices of order 6 have also been classified in \cite{nic}). A recent paper by Dita \cite{dita}
describes a  general construction which leads to parametric
families of complex Hadamard matrices in composite dimensions.
Another recent paper by Tadej and \.Zyczkowski \cite{karol} gives
an (admittedly incomplete) {\it catalogue} of complex Hadamard
matrices of {\it small order} (up to order 16).

The aim of this note is to show how tiling constructions of
Abelian groups can lead to constructions of complex Hadamard
matrices, and in this way to complement the catalogue of
\cite{karol} with new parametric families. In particular, we first
show how Dita's construction can be arrived at via a natural
tiling construction (this part does not lead to new results, but
it is an instructive example of how tiling and Hadamard matrices
are related). Second, we observe some regularities satisfied by
all Dita-type matrices, and thus arrive at an effective method to
decide whether a given complex Hadamard matrix is of Dita-type.
Then we use a combinatorial tiling construction due to Szab\'o
\cite{szabo} to produce Hadamard matrices {\it not of Dita-type},
and complement the catalogue of \cite{karol} with {\it new
parametric families} of order 8, 12 and 16.

\section{Recovering Dita's construction via tiling}\label{sec1}

This section describes a beautiful example of how seemingly
distant parts of mathematics are related to each other. A short
history of the construction is as follows.

Fuglede's conjecture states that a set in a locally compact
Abelian group (originally in $\RR^d$) is spectral (a notion to be
defined below) if and only if it tiles the group by translation.
(We remark that this conjecture has been disproved in dimensions 3
and higher \cite{mora, hadamard} but remains open in dimensions 1
and 2.) While tiling is a 'natural' notion, spectrality is less
so, and it is closely related to complex Hadamard matrices, as
explained below.  One approach to tackle the conjecture was to
look for 'canonical' {\it constructions for tiling Abelian groups}, and
see whether similar {\it constructions} work also {\it for spectral sets}.
This, indeed, turned out to be the case for a very general
construction (see Proposition \ref{prop1} below), and then this
general scheme of producing spectral sets {\it leads directly to Dita's
construction of complex Hadamard matrices}.

First, let us recall the most general form of Dita's construction,
formula (12) in \cite{dita} (his subsequent results on parametric
families of complex Hadamard matrices with some free parameters
follow easily from this formula, as described very well in
Proposition 3 and Theorem 2 of \cite{dita}).

\beql{ditaformula}
K:=\left [
\begin{array}{cccc}
m_{11}N_{1}&\cdot&\cdot&m_{1k}N_k\\
\cdot&\cdot&\cdot&\cdot\\
\cdot&\cdot&\cdot&\cdot\\
m_{k1}N_1&\cdot&\cdot&m_{kk}N_k
\end{array}
\right ] \eeq In this formula Dita assumes $m_{ij}$ to be the
entries of any $k\times k$ complex Hadamard matrix $M$, while
$N_j$ are any $n\times n$ complex Hadamard matrices (possibly
different from each other). Then he shows that $K$ is a complex
Hadamard matrix of order $kn$. While this construction seems
fairly natural, it may be remarkable that it has only been
discovered very recently \cite{dita} (we remark that an earlier,
less general construction was given in \cite{haagerup}), and that
it is so powerful that it leads to most of the parametric families
included in \cite{karol}.
\begin{definition}
A complex Hadamard matrix $K$ is called {\it Dita-type} if it is
equivalent to a matrix arising with formula \eqref{ditaformula}
(we use the standard notion of equivalence of Hadamard matrices
(see e.g. \cite{karol}), i.e. $K_1$ and $K_2$ are equivalent if
$K_1=D_1P_1K_2P_2D_2$ with unitary diagonal matrices $D_1, D_2$
and permutation matrices $P_1, P_2$.)
\end{definition}

Let us now turn to the definition of spectral sets and tiles, and
see how Dita's construction arises naturally via tiling of Abelian
groups.

\begin{definition}
Let $G$ be a locally compact Abelian group, and $\widehat G$ its
dual group (the group of characters). An open set $T\subset G$ is
said to be a translational tile if there is a disjoint union of some translated
copies of $T$ covering the whole group $G$ up to gaps of measure
zero (w.r.t Haar measure). $T\subset G$ is spectral if it has a
spectrum
 $S\subset \ft G$ such that the characters $\{\gamma|\gamma \in S\}$ restricted to $T$ form an
 orthogonal basis of $L^2(T)$. Then $(T,S)$ is called a spectral
 pair.
\end{definition}

{\it Remark 1.}  Let $\ZZ_N$ denote the cyclic group of $N$
elements. If $G$ is the Abelian group $\ZZ_N^d$, or $\mathbb Z^d$,
or $\TT^d$ then we identify elements of the group with column
vectors $\bf g$ of length $d$ (with entries $g_j=\frac{k}{N}$
($0\le k\le N-1$), or $g_j\in \ZZ$, or $g_j\in \TT$,
respectively). Also, we identify characters with row vectors $\bf
h$ of length $d$ (with entries $h_j=m$ ($0\le m\le N-1$), or
$h_j\in \TT$, or $h_j\in \ZZ$, respectively; it is also convenient
to identify $\TT$ with the interval $[0,1)$). The action of a
character is then described conveniently as
\beql{char}\gamma_h(g)=e^{2\pi i \langle \bf h, \bf g\rangle}\qquad \vh\in\ft G, \vg\in G.\eeq
These notations will be particularly useful to describe how spectral pairs lead to complex Hadamard matrices. Readers unfamiliar with this notation are advised to check the concrete numerical Example 1 in Section \ref{sec3}.

In the case of $G=\Z_N^d$ or $\Z^d$, if a finite set
$T=\{\vt_1,\dots,\vt_r\}\subset G$ has spectrum
$S=\{\vs_1,\dots,\vs_r\}\subset \ft G$ then by orthogonality
\beql{spectrum} \sum_{k=1}^re^{2\pi i\langle \vs_i-\vs_j,
\vt_k\rangle}=\sum_{k=1}^r\overline{\gamma}_{\vs_j}(\vt_k)\gamma_{\vs_i}(\vt_k)
=r \delta_{ij},\eeq so the matrix $[H]_{i,k}:=\left(e^{2\pi
i\langle \vs_i, \vt_k\rangle}\right)$ is an $r\times r$ {\it complex
Hadamard matrix} (i.e. a matrix with complex entries of absolute
value 1, such that the rows (and hence the columns) are
orthogonal). We call the matrix of exponents $[\log H]_{i,k}=\langle \vs_i,\vt_k
\rangle$ a {\it log-Hadamard matrix} (note that there is a factor $2\pi$  difference between \cite{nspec} and \cite{karol} as to the terminology 'log-Hadamard matrix'; here we adhere to the one used in \cite{nspec}). Finally, we have
arrived at the conclusion that $S$ is a spectrum of $T$ if and
only if the matrix product $ST$ is log-Hadamard. Accordingly, in
the case of $G=\TT^d$ we find it convenient to extend the
definition of spectrality to finite sets, too (finite sets are not
open and have measure zero in this case, so the original
definition is meaningless). $\hfill \square$
\begin{definition}
We say that a finite set $T\subset \TT^d$ is spectral if there
exists a set $S\subset \Z^d$ (as row vectors) such that $ST$ is
log-Hadamard.
\end{definition}

We now recall Proposition 2.2 from \cite{nspec} (the point is that
the analogous construction is {\it natural} for tiles (see
Proposition 2.1 in \cite{nspec}), and that is how this
construction was discovered for spectral sets).

\begin{proposition}\label{prop1}
Let $G$ be a finite Abelian group, and $H\le G$ a subgroup. Let
$T_1, T_2, \dots T_k\subset H$ be subsets of $H$ such that they
share a common spectrum in $\ft{H}$; i.e. there exists a set
$L\subset \ft{H}$ such that $L$ is a spectrum of $T_m$ for all
$1\le m\le k$. Consider any spectral pair $(Q,S)$ in the factor
group $G/H$, with $|Q|=k$, and take arbitrary representatives
$\vq_1, \vq_2, \dots \vq_k$ from the cosets of $H$ corresponding
to the set $Q$. Then the set $\C :=\cup_{m=1}^{k}(\vq_m+T_m)$ is
spectral in the group $G$.
\end{proposition}

\noindent{\it Proof.   } The proof is trivial, although the
notations are somewhat cumbersome. We will simply construct a
spectrum $\Sigma\subset \ft{G}$ for $\C$. Let $n$ denote the
number of elements in each $T_m$ (they necessarily have the same
number of elements as there exists a common spectrum), and
$\vt_r^m$ ($r=1, \dots n$ and $m=1, \dots k$) the $r$th element of
$T_m$. By assumption, there exist characters $\vl_j\in \ft{H}$
($j=1, \dots n$) such that the matrices $[A_m]_{j,r}:=[\vl_j
(\vt_r^m)]$ are $n\times n$ complex Hadamard for each $m$.
Let $\tilde \vl_j$ denote any extension of $\vl_j$ to a character
of $G$ (such extensions always exist, although not unique). Also,
the elements $\vs_1, \dots , \vs_k$ of $S\subset \ft{G/H}$ can be
identified with characters $\tilde \vs_i\in \ft{G}$ which are
constant on cosets of $H$. Then we consider the product characters
$\tilde \vs_i\tilde \vl_j$ and let $\Sigma :=\{\tilde \vs_i\tilde
\vl_j\}_{i,j}$ where $i=1, \dots , k$ and $j=1, \dots, n$. We
claim that $\Sigma$ is a spectrum of $\C$. For each $m=1, \dots k$
let $D_{L\vq_m}$ denote the $n\times n$ diagonal matrix with
entries $[D_{L\vq_m}]_{j,j}=\tilde \vl_j (\vq_m)$. Then, for fixed $i$ and $m$
the product characters $\tilde \vs_i\tilde \vl_j$ ($j=1, \dots ,
n$) restricted to the set $\vq_m+T_m=\{\vq_m+\vt_1^m, \dots ,
\vq_m+\vt_n^m\}$ simply give the $n\times n$ matrix
\beql{bim}B^{i,m}:=\tilde \vs_i (\vq_m) D_{Lq_m}A_m,\eeq
 because the entries
are given as $[B^{i,m}]_{j,r}=\tilde \vs_i\tilde \vl_j
(\vq_m+\vt_r^m)=\tilde \vs_i(\vq_m)\tilde\vl_j
(\vq_m)\tilde\vl_j(\vt_r^m)$. This means that the characters
$\tilde \vs_i\tilde \vl_j\in \Sigma$ restricted to $\C$ will give
the $nk\times nk$ block matrix
 \beql{sigmagamma} H:=\left [
\begin{array}{cccc}
B^{1,1}&\cdot&\cdot&B^{1,k}\\
\cdot&\cdot&\cdot&\cdot\\
\cdot&\cdot&\cdot&\cdot\\
B^{k,1}&\cdot&\cdot&B^{k,k}
\end{array}
\right ]. \eeq Now, observe that each block $B^{i,m}$ is given as a product
$\tilde \vs_i (\vq_m) D_{Lq_m}A_m$ where $N_m:=D_{Lq_m}A_m$ is a
complex Hadamard matrix (because $A_m$ is such and $D_{Lq_m}$ is a unitary diagonal matrix), and
$\tilde \vs_i (\vq_m)$ is the entry of a $k\times k$ complex
Hadamard matrix by the assumption that $S$ is a spectrum of $Q$.
Therefore $H$ is seen to be a complex Hadamard matrix arising
directly with formula \eqref{ditaformula}, and hence $\Sigma$ is
indeed a spectrum of $\C$. $\hfill \square$

{\it Remark 2.} We see that {\it the constructed spectral pair
$(\Sigma, \C)$ gives rise to a Dita-type matrix}. We remark,
however, that the set $\C$ might well have many other spectra than
the one constructed in the proof above (and other spectra might
produce complex Hadamard matrices not of the Dita-type). There is
no efficient algorithm known to list out all the spectra of a
given set. $\hfill \square$

{\it Remark 3.} The above Proposition was quoted verbatim from
\cite{nspec}, and remains in the finite group setting. This has
the disadvantage that the arising matrices are necessarily of the
Butson-type (i.e. containing roots of unity only), and one cannot
expect to obtain continuous parametric families of complex
Hadamard matrices.  However, the same construction works in the
infinite setting $G=\Z^d$ or $G=\TT^d$, too, and we now present
how {\it every Dita-type matrix arises in this manner}.

Assume that matrices $M$ and $N_m$ ($m=1, \dots , k$) are given,
and $K$ is constructed as in formula \eqref{ditaformula}. We aim
to recover $K$ with the construction of Proposition \ref{prop1}.

Let $G=\TT^d$ where $d=n+k$, and consider the subgroup $H_1=\TT^n$
(subgroup of vectors with last $k$ coordinates 0), and
$G/H_1=H_2=\TT^k$ (vectors with first $n$ coordinates 0). Then
$G=H_1\times H_2$.

Let $T_m=\log N_m$ denote the matrix of the exponents of the
entries of $N_m$, i.e. $[N_m]_{i,j}=e^{2\pi i [T_m]_{i,j}}$ (each
$T_m$ is defined $mod \  1$). Let $\tilde T_m\subset H_1$ denote
the set of vectors consisting of the columns of the log-Hadamard
matrix $T_m$ extended by 0's in the last $k$ coordinates. Then
each $\tilde T_m$ is spectral in $H_1$ and a common spectrum of
them is given by \beql{e1} E_1:=\left [
\begin{array}{ccccccc}
1&0&\dots&0&0&\dots &0\\
0&1&\dots&0&0&\dots &0\\
\vdots&&\ddots&\vdots&\vdots&&\vdots\\
0&0&\dots&1&0&\dots &0\\
\end{array}
\right].\eeq (This is because $E_1\tilde T_m=T_m$ is log-Hadamard
for each $m$.) Also, let $Q:=\log M$, and $\tilde \vq_j\in H_2$
denote the $j$th column of the log-Hadamard matrix $Q$ extended by
0's in the first $n$ coordinates. Then the set $\tilde Q=\{\tilde
\vq_1, \dots \tilde \vq_k\}\subset H_2$ is spectral in $H_2=G/H_1$
with spectrum \beql{e2} E_2:=\left [
\begin{array}{ccccccc}
0&0&\dots &0&1&\dots&0\\
\vdots&&&\vdots&\vdots&\ddots&\vdots\\
0&0&\dots &0&0&\dots&1
\end{array}
\right ]. \eeq (This is because $E_2\tilde Q=Q$ is log-Hadamard.)

As in Proposition \ref{prop1} above we define \beql{Gamma}
\C:=\cup_{m=1}^{k}(\vq_m+T_m)=\left [
\begin{array}{cccc|cccc|c|cccc}
&&&&&&&&&&&&\\
& T_1 && &  & T_2 && & \dots &  & T_k &&\\
&&&&&&&&&&&&\\
&&&&&&&&&&&&\\
\hline
|&|&\dots &|   &   |&|&\dots &|   &  \dots   &   |&|&\dots &|\\
\vq_1& \vq_1 &\dots &\vq_1   &   \vq_2& \vq_2 & \dots & \vq_2  & \dots  &\vq_k& \vq_k&\dots &\vq_k\\
|&|&\dots &|   &   |&|&\dots &|   &  \dots   &   |&|&\dots &|
\end{array}
\right ] \eeq Then, the spectrum $\Sigma$ constructed in the proof
of Proposition \ref{prop1} takes the form '$\Sigma =E_1+E_2$',
i.e. \beql{Sigma} \Sigma:=\left [
\begin{array}{cccc|cccc}
1&0&\dots&0 &1&0&\dots&0\\
0&1&\dots&0 &1&0&\dots&0\\
\vdots&&\ddots&\vdots&\vdots&\vdots\\
0&0&\dots&1 &1&0&\dots&0\\
\hline
1&0&\dots&0 &0&1&\dots&0\\
0&1&\dots&0 &0&1&\dots&0\\
\vdots&&\ddots&\vdots&\vdots&\vdots\\
0&0&\dots&1 &0&1&\dots&0\\
\hline
\vdots &&& & \vdots &&\\
\hline
1&0&\dots&0 &0&0&\dots&1\\
0&1&\dots&0 &0&0&\dots&1\\
\vdots&&\ddots&\vdots&\vdots&\vdots\\
0&0&\dots&1 &0&0&\dots&1
\end{array}
\right ] \eeq Finally, the $nk\times nk$ log-Hadamard matrix
arising from the spectral pair $( \C , \Sigma )$ is the product
$\Sigma\C$, and it gives back exactly the log-Hadamard matrix
$\log K$, as desired.

\beql{szorzat} \Sigma\C= \log K=\left [
\begin{array}{cccc}
q_{11}+T_{1}&q_{12}+T_{2}&\dots&q_{1k}+T_k\\
q_{21}+T_{1}&q_{22}+T_{2}&\dots&q_{2k}+T_k\\
\vdots&\vdots&&\vdots\\
q_{k1}+T_{1}&q_{k2}+T_{2}&\dots&q_{kk}+T_k\end{array} \right ]
\eeq
$\hfill \square$
\section{Other tiling constructions yielding new families of complex Hadamard matrices }\label{sec3}

Once the connection between tiling and complex Hadamard matrices
has been noticed, it is natural to look for tiling constructions
other than that of Proposition \ref{prop1} above, in the hope of
producing new complex Hadamard matrices not of the
Dita-type. Furthermore, when a new complex Hadamard matrix $M$ is
discovered, the 'linear variation of phases' method of
\cite{karol} gives hope to find  new parametric affine families of
complex Hadamard matrices stemming from $M$. This is exactly the
route we are going to follow in this section. First, we show how {\it a
tiling method of Szab\'o \cite{szabo} leads to complex Hadamard
matrices not of the Dita-type}. Then, {\it stemming from these matrices},
we produce {\it new parametric families of order 8, 12, and 16} which have
not been present in the literature so far and which complement the
recent catalogue \cite{karol}.

It turns out that the (tiling analogue of) the construction of
Proposition \ref{prop1} is so general that it is not trivial to
produce tilings which do not arise in such manner. In fact, it was
once asked by Sands \cite{sands} whether {\it every} tiling of
finite Abelian groups is such that one of the factors is contained
in a subgroup (note that such tilings correspond to the special
case $Q=G/H$ in the tiling analogue of Proposition \ref{prop1}).
This question was then answered in the negative by {\it a construction
of Szab\'o} \cite{szabo}, which we now turn to.

Assume $G=\Z_{p_1q_1}\times \Z_{p_2q_2}\times \Z_{p_3q_3}$ where
$p_j, q_j\ge 2$. The idea of Szab\'o is to take the obvious
tiling $G=A+B$ where
\beql{a}A=\{0, \frac{1}{p_1q_1}, \dots \frac{p_1-1}{p_1q_1}\}\times \{0, \frac{1}{p_2q_2}, \dots
\frac{p_2-1}{p_2q_2}\}\times \{0, \frac{1}{p_3q_3}, \dots \frac{p_3-1}{p_3q_3}\}\eeq
 and $B=\{0, \frac{1}{q_1}, \frac{2}{q_1}, \dots
\frac{q_1-1}{q_1}\}\times \{0, \frac{1}{q_2}, \frac{2}{q_2}, \dots \frac{q_2-1}{q_2}\}\times \{0, \frac{1}{q_3},
\frac{2}{q_3}, \dots \frac{q_3-1}{q_3}\}$ and then modify the grid $B$ by pushing
three grid-lines in different directions (see \cite{szabo} for
details; we do not describe the details here as we do not directly use this construction in this paper, it serves only as a guide to our spectral analogue below). Here we use the {\it analogous construction for spectral
sets} which we now describe in detail (it may be easier to follow the general construction by
looking at the specific Example 1 below).

Consider the set $A$ above. By formula \eqref{spectrum} a set
$S\subset \ft{G}$ is a spectrum of $A$ if and only if $|S|=|A|$
and $S-S\subset Z_A\cup \{0\}:=\{\vr\in\ft{G}: \ft \chi_A (\vr
)=0\}\cup \{0\}$ ($\chi_A$ denotes the indicator function of $A$, and the Fourier transform $\ft \chi_A$ is evaluated at some $\vr\in \ft{G}$ as $\ft \chi_A (\vr )=\sum_{\va\in A}e^{2\pi i \langle \vr, \va \rangle}$).
For a more detailed discussion of this fact see e.g.
\cite{nspec}. Recall that $\ft{G}$ is identified with
3-dimensional row vectors. It is clear that if $\vr=(r_1, r_2,
r_3)\in \ft{G}$ is such that $q_1$ divides $r_1$ and $r_1\ne 0$ then $\ft \chi_A (\vr
)=0$ (all sub-sums become 0 with fixing the second and third
coordinate and letting the first one vary in $A$). Similarly, if
$q_2|r_2\ne 0$ or $q_3|r_3\ne 0$ then $\ft \chi_A (\vr )=0$.
Therefore the grid
\beql{s}S=\{0, q_1, \dots (p_1-1)q_1\}\times \{0, q_2,
\dots (p_2-1)q_2\}\times \{0, q_3, \dots (p_3-1)q_3\}\eeq
is a
spectrum of $A$. Using an analogous idea to that of Szab\'o we now modify this
grid.

Consider the grid-line $L_1:=\{\{0, q_1, \dots (p_1-1)q_1\}\times
\{q_2\}\times \{0\}$ and change it to $L_1':=\{1, q_1+1, \dots
(p_1-1)q_1+1\}\times \{q_2\}\times \{0\}$ (adding +1 to the first
coordinates). Similarly, change $L_2:=\{0\}\times \{0, q_2, \dots
(p_2-1)q_2\}\times\{q_3\}$ to $L_2':=\{0\}\times \{1, q_2+1, \dots
(p_2-1)q_2+1\}\times\{q_3\}$, and change $L_3:=\{q_1\}\times
\{0\}\times \{0, q_3, \dots (p_3-1)q_3\}$ to $L_3':=\{q_1\}\times
\{0\}\times \{1, q_3+1, \dots (p_3-1)q_3+1\}$. It is easy to see
that
\beql{s'}S':=S\cup (L_1'\cup L_2'\cup L_3') \setminus (L_1\cup
L_2\cup L_3)\eeq
is still a spectrum of $A$. Indeed, for any $\vr\in
S'-S'$ it still holds that either the first coordinate is
divisible by $q_1$ or the second by $q_2$ or the third by $q_3$.
Then the spectral pair $(A, S')$ gives rise to a complex Hadamard
matrix of size $p_1p_2p_3$.
Below we will
apply this construction in the groups $G_1=\Z_{2\cdot 2}\times
\Z_{2\cdot 2}\times \Z_{2\cdot 2}$, $G_2=\Z_{2\cdot 2}\times
\Z_{2\cdot 2}\times \Z_{3\cdot 3}$ and $G_3=\Z_{2\cdot 2}\times
\Z_{4\cdot 2}\times \Z_{2\cdot 4}$ (it may be instructive to see the step-by-step numerical exposition of the construction in Example 1 in group $G_1$ below).

We will then prove that these matrices are {\it not of the
Dita-type}. (It would be very interesting to see a proof of a
general statement that all matrices arising with the above
construction are non-Dita-type.) As a result we will conclude that these matrices have not been included in the
catalogue \cite{karol}.

\noindent{\it Remark 4.} We can see from the construction above that the {\it size} of the arising matrix is $p_1p_2p_3$, while the numbers $q_1, q_2, q_3$ are chosen arbitrarily to determine the group we are working in. It is not clear whether different choices of $q_1, q_2, q_3$ lead to non-equivalent Hadamard matrices. In this paper we only list the three examples for which the dimension is not greater than 16 (as in \cite{karol}) and for which we can {\it prove} that the arising matrices are new, i.e. non-equivalent to any matrix listed in \cite{karol}. $\hfill \square$

\noindent{\bf Example 1.} Let us follow the construction above, step by step, in $G_1=\Z_{2\cdot 2}\times
\Z_{2\cdot 2}\times \Z_{2\cdot 2}=\Z_4\times \Z_4\times \Z_4$.

By \eqref{a} we take $A=\{0, \frac{1}{4}\}\times \{0, \frac{1}{4}\}\times \{0, \frac{1}{4}\}$. This is a Cartesian product, each element of which is a 3-dimensional vector composed of 0's and $\frac{1}{4}$'s. We list out the elements in lexicographical order as
\beql{amatrix}
A=\frac{1}{4}\left[\begin{array}{cccccccc}
0 & 0 & 0 & 0 & 1 & 1 & 1 & 1 \\
 0 & 0 & 1 & 1 & 0 & 0 & 1 & 1 \\
 0 & 1 & 0 & 1 & 0 & 1 & 0 & 1
\end{array}\right],
\eeq where the columns represent the elements of $A\subset G_1$, in accordance with our notation introduced earlier. (The order of the elements is up to our choice, but a permutation of the elements only corresponds to a permutation of the columns of the matrix $S_8$ below.)

Then, by equation \eqref{s} we have $S=\{0,2\}\times \{0, 2\}\times \{0,2\}$, which we list out (also in lexicographical order) as
\beql{smatrix}
S=\left[\begin{array}{ccc}
0 & 0 & 0\\
0 & 0 & 2\\
0 & 2 & 0\\
0 & 2 & 2\\
2 & 0 & 0\\
2 & 0 & 2\\
2 & 2 & 0\\
2 & 2 & 2\\
\end{array}\right]
\eeq
Now, $S$ is a spectrum of $A$, therefore the product $SA$ already gives a log-Hadamard matrix but we do not take that matrix (which {\it is Dita-type}, as can be verified by the reader), but modify the set $S$ first. The grid-line $L_1$ in $S$ is given as $L_1=\{0,2\}\times \{2\}\times \{0\}=\{(0,2,0); (2,2,0)\}$. This we replace by $L_1'=\{(1,2,0); (3,2,0)\}$. Similarly, the grid-line $L_2=\{(0,0,2);(0,2,2)\}$ is replaced by $L_2'=\{(0,1,2);(0,3,2)\}$ and finally $L_3=\{(2,0,0);(2,0,2)\}$ by $L_3'=\{(2,0,1);(2,0,3)\}$. Therefore, by \eqref{s'} we get
\beql{s'matrix}
S'=S\cup (L_1'\cup L_2'\cup L_3') \setminus (L_1\cup
L_2\cup L_3)=
\left[\begin{array}{ccc}
 0 & 0 & 0 \\
 0 & 1 & 2 \\
 0 & 3 & 2 \\
 1 & 2 & 0 \\
 2 & 0 & 1 \\
 2 & 0 & 3 \\
 2 & 2 & 2 \\
 3 & 2 & 0
\end{array}\right]
\eeq (Once again, the order of the elements of $S'$ is arbitrary, and we take lexicographical order.)
The point is, as explained above in the general description of this construction, that the set $S'$ is still a spectrum of $A$. Therefore the matrix product $S'A$ is a log-Hadamard matrix (we reduce the entries $mod \ 1$ because the integer part of an entry plays no role after exponentiation; e.g. $\frac{5}{4}\equiv \frac{1}{4}$) given by:
\beql{logs8} S'A=\log S_8=
\frac{1}{4}\left[\begin{array}{cccccccc}
 0 & 0 & 0 & 0 & 0 & 0 & 0 & 0 \\
 0 & 2 & 1 & 3 & 0 & 2 & 1 & 3 \\
 0 & 2 & 3 & 1 & 0 & 2 & 3 & 1 \\
 0 & 0 & 2 & 2 & 1 & 1 & 3 & 3 \\
 0 & 1 & 0 & 1 & 2 & 3 & 2 & 3 \\
 0 & 3 & 0 & 3 & 2 & 1 & 2 & 1 \\
 0 & 2 & 2 & 0 & 2 & 0 & 0 & 2 \\
 0 & 0 & 2 & 2 & 3 & 3 & 1 & 1
\end{array}\right]\eeq with the corresponding Hadamard matrix given by
\beql{s8}S_8=
\left[\begin{array}{rrrrrrrr}
 1 & 1 & 1 & 1 & 1 & 1 & 1 & 1 \\
 1 & -1 & \textbf{i} & -\textbf{i} & 1 & -1 & \textbf{i} & -\textbf{i} \\
 1 & -1 & -\textbf{i} & \textbf{i} & 1 & -1 & -\textbf{i} & \textbf{i} \\
 1 & 1 & -1 & -1 & \textbf{i} & \textbf{i} & -\textbf{i} & -\textbf{i} \\
 1 & \textbf{i} & 1 & \textbf{i} & -1 & -\textbf{i} & -1 & -\textbf{i} \\
 1 & -\textbf{i} & 1 & -\textbf{i} & -1 & \textbf{i} & -1 & \textbf{i} \\
 1 & -1 & -1 & 1 & -1 & 1 & 1 & -1 \\
 1 & 1 & -1 & -1 & -\textbf{i} & -\textbf{i} & \textbf{i} & \textbf{i}
\end{array}\right]
\eeq$\hfill \square$

Having described {\it how to produce} the matrix $S_8$ the remaining questions are whether $S_8$ is {\it new (i.e. not already included in the catalogue \cite{karol})}, and whether {\it any parametric family of complex Hadamard matrices} stems from $S_8$.

We will first proceed to show that $S_8$ is {\it not Dita-type}
(nor is it its transpose). This is a delicate matter, as not many
criteria are known to decide inequivalence of Hadamard matrices.
The Haagerup condition with the invariant set $\Lambda := \{
h_{ij}\overline h_{kj}h_{kl}\overline h_{il}\}$ (see
\cite{haagerup} and Lemma 2.5 in \cite{karol}) cannot be used
here. Also, the elegant characterization of equivalence classes of
Kronecker products of Fourier matrices \cite{tadej} does not apply
to $S_8$. The 'regular' structure of a Dita-type matrix must be
exploited in some way. The key observation relies on the following

\begin{definition}
Let $L$ be an $N\times N$ real matrix. For an index set
$I=\{i_1,i_2,\dots, i_n\} \subset \{1,2,\dots, N\}$ two rows (or
columns) $\vs$ and $\vq$ are called {\it $I$-equivalent}, in
notation $\vs\sim_I\vq$, if the fractional part of the entry-wise
differences $s_i-q_i$ are the same for every $i\in I$ (we need to
consider fractional parts as the entries of a log-Hadamard matrix
are defined only $mod \ 1$). Two rows (or columns) $\vs$ and $\vq$
are called $(d)$-$n$-equivalent if there exist $n$-element
disjoint sets of indices $I_1, \dots , I_d$ such that
$\vs\sim_{I_j}\vq$ for all $j=1, \dots , d$.
 \end{definition}

We have the following trivial observation.
\begin{proposition}
Let $L$ be an $N\times N$ complex Hadamard matrix. Assume that there exist an index set  $I=\{i_1,i_2,\dots, i_n\} \subset \{1,2,\dots, N\}$ and $m$ different rows (resp. columns) $\vr_{s_1}, \dots \vr_{s_m}$ in the log-Hadamard matrix $\log L$ such that each two of them are $I$-equivalent. Let $M$ be any complex Hadamard matrix equivalent to $L$. Then the same property holds for $\log M$, i.e. there exist an index set $J=\{j_1,j_2,\dots, j_n\} \subset \{1,2,\dots, N\}$ and $m$ different rows (resp. columns) $\vr_{k_1}, \dots \vr_{k_m}$ such that each two of them are $J$-equivalent. (Of course, the index sets $I$ and $\{s_1, \dots s_m\}$ might not be the same as $J$ and $\{k_1, \dots k_m\}$.)
\end{proposition}
\noindent{\it Proof.}
It follows from the definition of the equivalence of Hadamard matrices that $\log M$ is obtained from $\log L$ by permutation of rows and columns, and addition of constants to rows and columns. It is clear that such operations preserve the existing equivalences between rows and columns (with the index sets being altered according to the permutations used).
$\hfill \square$

The essence of the proposition is that {\it "existing equivalences between rows and columns are retained"}. The next main point is that there are many equivalences among the rows of a Dita-type matrix and we will see that such equivalences are not present in $\log S_8$.

By formula \eqref{ditaformula}, the structure of an $N\times N$
Dita-type matrix $D$ (where $N=nk$) implies for the log-Hadamard
matrix $\log D$ that there exists a partition of indices to
$n$-element sets $I_1=\{1,2,\dots n\}, \dots , I_k=\{(k-1)n+1, \dots kn\}$ and $k$-tuples of rows
$R_j=\{\vr_{j}, \vr_{j+n} \dots \vr_{j+(k-1)n}\}$ ($j=1, \dots n$) such that any
two rows in a fixed $k$-tuple are equivalent with respect to any
of the $I_m$'s, i.e. $\vr_{j+(i-1)n}\sim_{I_m} \vr_{j+(s-1)n}$ for all $j=1,
\dots n$, and $i,s,m=1, \dots k$. In other words, in any $k$-tuple
$R_j$ any two rows are $(k)$-$n$-equivalent with respect to the
$I_m$'s. We will use the terminology {\it $(k)$-$n$-Dita-type} for such matrices $D$. Naturally, the same property holds for the transposed of a
$(k)$-$n$-Dita-type matrix, with the role of rows and columns interchanged.

This observation makes it possible to prove the following
\begin{proposition}
 $S_8$ and its
transposed are not Dita-type.\end{proposition}
\noindent{\it Proof.} The matrix size being $8\times 8$ the only
possible values for $n$ are 2 and 4 (with $k$ being 4 and 2,
respectively). Therefore we only need to check existing (2)-4-equivalences and (4)-2-equivalences in $\log S_8$ and its transposed.

First, let us assume that $n=4, k=2$ and look for (2)-4-equivalences among the rows of $\log S_8$.
If $S_8$ were (2)-4-Dita type,
there should be a partition of indices to two 4-element sets $I_1, I_2$
such that in $\log S_8$ four pairs of rows are equivalent with
respect to $I_1$, $I_2$. The first row $\vr_1$ of $\log S_8$
consists of zeros only, therefore it must be paired with a row
containing only two different values. There is only one such row $\vr_7$ and then the index sets must correspond to the position of 0's and 2's in $\vr_7$, i.e. $I_1=\{1,4,6,7\}$ and $I_2=\{2,3,5,8\}$. However, there should exist {\it three further pairs
of rows} which are equivalent with respect to the same set of
indices $I_1$, $I_2$. It is easy to check that such pairs do not exist (e.g. the second row $\vr_2$ is not (2)-4-equivalent with respect to $I_1, I_2$ to any other row), and hence $S_8$ cannot be (2)-4-Dita type.

To check the transposed
matrix we interchange the role of rows and columns and see that
the first column $\vc_1$ of $\log S_8$ (all zeros) should be paired with a column containing two values only. But such column does not exist, therefore $\vc_1$ is not (2)-4-equivalent to any other column, and hence the transposed of $S_8$ cannot be (2)-4-Dita type.

Let us turn to the case $n=2, k=4$. If $S_8$ were (4)-2-Dita type,
there should be a partition of indices to four 2-element sets $I_1, I_2, I_3, I_4$
such that in $\log S_8$ two 4-tuples of rows $R_1=\{\vr_{s_1}, \dots , \vr_{s_4}\}$ and $R_2=\{\vr_{s_5}, \dots , \vr_{s_8}\}$ are equivalent with
respect to $I_1, I_2, I_3, I_4$. Assume, without loss of generality that $1\in I_1$ (i.e. $I_1=\{1, m\}$ for some $m$) and that $\vr_{s_1}=\vr_1$. Then $\vr_{s_2}, \vr_{s_3}, \vr_{s_4}$ are $I_1$-equivalent to $\vr_1$ which implies that there should be a $4\times 2$ block of 0's in $\log S_8$ corresponding to $R_1$ and $I_1$, i.e. $[\log S_8]_{i,j}=0$ for all $i\in R_1$ and $j\in I_1$. Such block of 0's does not exist, therefore $S_8$ is not (4)-2-Dita-type.

In the transposed case there exists such a $2\times 4$ block of zeros, corresponding to the row indices $I_1=\{1,7\}$ and column indices $C_1=\{1, 4,6,7\}$. This means that there should be further two-element index sets $I_2, I_3, I_4$ such that the columns $\{c_1, c_4, c_6, c_7\}$ are equivalent with respect to $I_2, I_3, I_4$. It is trivial to check that such indices do not exist.
This concludes the proof that $S_8$ and its transposed
are not Dita-type. $\hfill \square$

The significance of this fact is that the {\it only known $8\times 8$
parametric family of complex Hadamard matrices so far is the one
constructed by Dita's method} (see \cite{karol}). It is an affine
family $F_8^{(5)}(a,b,c,d,e)$ containing 5 free parameters. We have established that this
family does not go through $S_8$, therefore $S_8$ is indeed {\it new}. In particular, the matrix $S_8$ cannot be equivalent to any of the
well-known tensor products of Fourier-matrices $F_2\otimes F_2\otimes F_2$, $F_4\otimes F_2$, $F_8$ which are all contained in the family $F_8^{(5)}(a,b,c,d,e)$.

Now, applying to $S_8$ the
linear variation of phases method of \cite{karol} one can hope to
obtain new parametric families of complex Hadamard matrices.
Indeed, we have been able to obtain\footnote{The authors are
grateful to W. Tadej who extended the 3-parameter family
$S_8(a,b,c)$ communicated to him.} the following
 maximal affine 4-parameter family (the notation is used as in \cite{karol}, i.e. the symbol $\circ$ denotes the Hadamard product of two matrices $[H_1\circ H_2]_{i,j}=[H_1]_{i,j}\cdot [H_2]_{i,j}$, and the symbol EXP denotes the entrywise exponential operation $[EXP \ H]_{i,j}=exp ([H]_{i,j})$):
$S^{(4)}_8(a,b,c,d)=S_8\circ EXP (\textbf{i}R^{(4)}_8(a,b,c,d)$, where
\beql{r8}R^{(4)}_8(a,b,c,d)=
\left[\begin{array}{cccccccc}
 \bullet & \bullet & \bullet & \bullet & \bullet & \bullet & \bullet & \bullet \\
 \bullet & d & a & a-d & d & \bullet & a-d & a \\
 \bullet & d & a & a-d & d & \bullet & a-d & a \\
 \bullet & d & d & \bullet & b & b-d & b-d & b \\
 \bullet & c & d & c-d & d & c-d & \bullet & c \\
 \bullet & c & d & c-d & d & c-d & \bullet & c \\
 \bullet & \bullet & \bullet & \bullet & \bullet & \bullet & \bullet & \bullet \\
 \bullet & d & d & \bullet & b & b-d & b-d & b
\end{array}\right]\eeq
We do not claim that each matrix in $S^{(4)}_8(a,b,c,d)$ is
non-Dita-type (in fact, it is not hard to see that the orbit
$S^{(4)}_8(a,b,c,d)$ contains the only real $8\times 8$ Hadamard matrix $H_8$,
which {\it is} Dita-type, so the families $F_8^{(5)}(a,b,c,d,e)$ and $S^{(4)}_8(a,b,c,d)$ {\it intersect} each other at $H_8$). However, this is certainly true in a small
neighbourhood of $S_8$ as the set of Dita-matrices is closed.

\noindent{\bf Example 2.}
We now turn to $N=16$ and the group $G_3=\Z_{2\cdot 2}\times
\Z_{4\cdot 2}\times \Z_{2\cdot 4}$ (we leave $N=12$ last, as the
discussion is slightly different there).

The construction described above yields the following
matrices:
\beql{a16}
A_{G_3}=\frac{1}{8}\left[\begin{array}{cccccccccccccccc}
 0 & 0 & 0 & 0 & 0 & 0 & 0 & 0 & 2 & 2 & 2 & 2 & 2 & 2 & 2 & 2 \\
 0 & 0 & 1 & 1 & 2 & 2 & 3 & 3 & 0 & 0 & 1 & 1 & 2 & 2 & 3 & 3 \\
 0 & 1 & 0 & 1 & 0 & 1 & 0 & 1 & 0 & 1 & 0 & 1 & 0 & 1 & 0 & 1
\end{array}\right],
\eeq
and we give $S'_{G_3}$ in transposed layout to save space
\beql{s'16}
\left(S'_{G_3}\right)^T=\left[
\begin{array}{cccccccccccccccc}
 0 & 0 & 0 & 0 & 0 & 0 & 0 & 1 & 2 & 2 & 2 & 2 & 2 & 2 & 2 & 3 \\
 0 & 1 & 3 & 4 & 5 & 6 & 7 & 2 & 0 & 0 & 2 & 4 & 4 & 6 & 6 & 2 \\
 0 & 4 & 4 & 0 & 4 & 0 & 4 & 0 & 1 & 5 & 4 & 0 & 4 & 0 & 4 & 0
\end{array}
\right],
\eeq
and the arising log-Hadamard matrix (containing 8th roots of unity):

\beql{s16}
S'_{G_3}A_{G_3}=\log S_{16}=\frac{1}{8}\left[
\begin{array}{cccccccccccccccc}
 0 & 0 & 0 & 0 & 0 & 0 & 0 & 0 & 0 & 0 & 0 & 0 & 0 & 0 & 0 & 0 \\
 0 & 4 & 1 & 5 & 2 & 6 & 3 & 7 & 0 & 4 & 1 & 5 & 2 & 6 & 3 & 7 \\
 0 & 4 & 3 & 7 & 6 & 2 & 1 & 5 & 0 & 4 & 3 & 7 & 6 & 2 & 1 & 5 \\
 0 & 0 & 4 & 4 & 0 & 0 & 4 & 4 & 0 & 0 & 4 & 4 & 0 & 0 & 4 & 4 \\
 0 & 4 & 5 & 1 & 2 & 6 & 7 & 3 & 0 & 4 & 5 & 1 & 2 & 6 & 7 & 3 \\
 0 & 0 & 6 & 6 & 4 & 4 & 2 & 2 & 0 & 0 & 6 & 6 & 4 & 4 & 2 & 2 \\
 0 & 4 & 7 & 3 & 6 & 2 & 5 & 1 & 0 & 4 & 7 & 3 & 6 & 2 & 5 & 1 \\
 0 & 0 & 2 & 2 & 4 & 4 & 6 & 6 & 2 & 2 & 4 & 4 & 6 & 6 & 0 & 0 \\
 0 & 1 & 0 & 1 & 0 & 1 & 0 & 1 & 4 & 5 & 4 & 5 & 4 & 5 & 4 & 5 \\
 0 & 5 & 0 & 5 & 0 & 5 & 0 & 5 & 4 & 1 & 4 & 1 & 4 & 1 & 4 & 1 \\
 0 & 4 & 2 & 6 & 4 & 0 & 6 & 2 & 4 & 0 & 6 & 2 & 0 & 4 & 2 & 6 \\
 0 & 0 & 4 & 4 & 0 & 0 & 4 & 4 & 4 & 4 & 0 & 0 & 4 & 4 & 0 & 0 \\
 0 & 4 & 4 & 0 & 0 & 4 & 4 & 0 & 4 & 0 & 0 & 4 & 4 & 0 & 0 & 4 \\
 0 & 0 & 6 & 6 & 4 & 4 & 2 & 2 & 4 & 4 & 2 & 2 & 0 & 0 & 6 & 6 \\
 0 & 4 & 6 & 2 & 4 & 0 & 2 & 6 & 4 & 0 & 2 & 6 & 0 & 4 & 6 & 2 \\
 0 & 0 & 2 & 2 & 4 & 4 & 6 & 6 & 6 & 6 & 0 & 0 & 2 & 2 & 4 & 4
\end{array}\right]\eeq $\hfill \square$

\begin{proposition}
$S_{16}$ and its transposed are not Dita-type.
\end{proposition}
\noindent{\it Proof.} By checking existing $I$-equivalences between rows (and columns)
it is elementary (but tedious) to show that $S_{16}$ (and its
transposed) is {\it not Dita-type}. To find possible index sets $I$ and $I$-equivalences between rows (resp. columns, in the transposed case) it is perhaps most convenient to note the position of 0's in $\log S_{16}$ and look for $2\times 8$, $4\times 4$ and $8\times 2$ blocks of 0's as in the last part of the proof concerning $S_8$. Then each of these $I$-patterns can be excluded by looking at further rows (resp. columns).$\hfill \square$

The significance of this fact, once
again, is that {\it the only known $16\times 16$ parametric family so far is
the one constructed with Dita's method} (see \cite{karol}). It is an affine
family $F_{16}^{(17)}(a,b,c,d,e,f,g,h,i,j,k,l,m,n,o,p,r)$ containing 17 free parameters. We have established that this
family does not go through $S_{16}$. In particular, the matrix $S_{16}$ cannot be equivalent to any of the
well-known tensor products of Fourier-matrices $F_2\otimes F_2\otimes F_2\otimes F_2, F_4\otimes F_2\otimes F_2, F_4\otimes F_4, F_8\otimes F_2, F_{16}$ which are all contained in the family $F_{16}^{(17)}$.

By applying the linear
variation of phases method of \cite{karol} we have been able to
find the following 11-parameter affine family stemming from
$S_{16}$. Again, we can claim that the members of this family are
not Dita-type in a neighbourhood of $S_{16}$. However, in this
case we do not know whether this affine family is maximal or
further parameters can be introduced.

\beql{s16a}S^{(11)}_{16}(a,b,c,d,e,f,g,h,i,j,k)=S_{16}\circ EXP(\textbf{i} R^{(11)}_{16}(a,b,c,d,e,f,g,h,i,j,k)),  \ \ \ \mathrm{where}\eeq
\pagebreak
\beql{r16}  R^{(11)}_{16}(a,b,c,d,e,f,g,h,i,j,k)=\eeq
\[\left[\begin{array}{cccccccccccccccc}
 \bullet & \bullet & \bullet & \bullet & \bullet & \bullet & \bullet & \bullet & \bullet & \bullet & \bullet & \bullet & \bullet & \bullet & \bullet & \bullet \\
 \bullet & \bullet & b & b & d & d & b+j & b+j & \bullet & \bullet & b & b & d & d & b+j & b+j \\
 \bullet & \bullet & c & c & d & d & c+j & c+j & \bullet & \bullet & c & c & d & d & c+j & c+j \\
 \bullet & \bullet & \bullet & \bullet & \bullet & \bullet & \bullet & \bullet & \bullet & \bullet & \bullet & \bullet & \bullet & \bullet & \bullet & \bullet \\
 \bullet & \bullet & b & b & d & d & b+j & b+j & \bullet & \bullet & b & b & d & d & b+j & b+j \\
 \bullet & h & h+i & i & \bullet & h & h+i & i & h & \bullet & i & h+i & h & \bullet & i & h+i \\
 \bullet & \bullet & c & c & d & d & c+j & c+j & \bullet & \bullet & c & c & d & d & c+j & c+j \\
 \bullet & \bullet & i & i & \bullet & \bullet & i & i & g & g & g+i & g+i & g & g & g+i & g+i \\
 \bullet & a & k & a & \bullet & a & k & a & \bullet & a & k & a & \bullet & a & k & a \\
 \bullet & a & k & a & \bullet & a & k & a & \bullet & a & k & a & \bullet & a & k & a \\
 \bullet & h & h+i & i & \bullet & h & h+i & i & h & \bullet & i & h+i & h & \bullet & i & h+i \\
 \bullet & f & k & f & \bullet & f & k & f & \bullet & f & k & f & \bullet & f & k & f \\
 \bullet & f & k & f & \bullet & f & k & f & \bullet & f & k & f & \bullet & f & k & f \\
 \bullet & e & i & e+i & \bullet & e & i & e+i & \bullet & e & i & e+i & \bullet & e & i & e+i \\
 \bullet & e & i & e+i & \bullet & e & i & e+i & \bullet & e & i & e+i & \bullet & e & i & e+i \\
 \bullet & \bullet & i & i & \bullet & \bullet & i & i & g & g & g+i & g+i & g & g & g+i & g+i
\end{array}\right]\]

\noindent{\bf Example 3.}
Finally, we turn to the case $N=12$ and the group $G_2=\Z_{2\cdot
2}\times \Z_{2\cdot 2}\times \Z_{3\cdot 3}$. Here our
construction yields the following matrices:
\beql{a12}
A_{G_2}=\frac{1}{36}\left[\begin{array}{cccccccccccc}
 0 & 0 & 0 & 0 & 0 & 0 & 9 & 9 & 9 & 9 & 9 & 9 \\
 0 & 0 & 0 & 9 & 9 & 9 & 0 & 0 & 0 & 9 & 9 & 9 \\
 0 & 4 & 8 & 0 & 4 & 8 & 0 & 4 & 8 & 0 & 4 & 8
\end{array}\right]
\eeq
and $S'_{G_2}$ in transposed layout
\beql{s'12}
\left(S'_{G_2}\right)^T=\left[
\begin{array}{cccccccccccc}
 0 & 0 & 0 & 0 & 0 & 1 & 2 & 2 & 2 & 2 & 2 & 3 \\
 0 & 0 & 1 & 2 & 3 & 2 & 0 & 0 & 0 & 2 & 2 & 2 \\
 0 & 6 & 3 & 6 & 3 & 0 & 1 & 4 & 7 & 3 & 6 & 0
\end{array}
\right]
\eeq

and the arising log-Hadamard matrix (containing 36th roots of unity):

\beql{s12}\log S_{12}=
\frac{1}{36}\left[\begin{array}{cccccccccccc}
 0 & 0 & 0 & 0 & 0 & 0 & 0 & 0 & 0 & 0 & 0 & 0 \\
 0 & 24 & 12 & 0 & 24 & 12 & 0 & 24 & 12 & 0 & 24 & 12 \\
 0 & 12 & 24 & 9 & 21 & 33 & 0 & 12 & 24 & 9 & 21 & 33 \\
 0 & 24 & 12 & 18 & 6 & 30 & 0 & 24 & 12 & 18 & 6 & 30 \\
 0 & 12 & 24 & 27 & 3 & 15 & 0 & 12 & 24 & 27 & 3 & 15 \\
 0 & 0 & 0 & 18 & 18 & 18 & 9 & 9 & 9 & 27 & 27 & 27 \\
 0 & 4 & 8 & 0 & 4 & 8 & 18 & 22 & 26 & 18 & 22 & 26 \\
 0 & 16 & 32 & 0 & 16 & 32 & 18 & 34 & 14 & 18 & 34 & 14 \\
 0 & 28 & 20 & 0 & 28 & 20 & 18 & 10 & 2 & 18 & 10 & 2 \\
 0 & 12 & 24 & 18 & 30 & 6 & 18 & 30 & 6 & 0 & 12 & 24 \\
 0 & 24 & 12 & 18 & 6 & 30 & 18 & 6 & 30 & 0 & 24 & 12 \\
 0 & 0 & 0 & 18 & 18 & 18 & 27 & 27 & 27 & 9 & 9 & 9
\end{array}\right]\eeq
$\hfill \square$

The difference in the discussion of this case lies in the fact
that there are several parametric families known already for
$N=12$. The catalogue \cite{karol} lists seven 9-parameter
families stemming from $F_{12}$, and only one of them is certain
to be constructed with Dita's method. (We remark that possible
permutational equivalences between these families are still
unclear.) Also, there are other $12\times 12$ families listed in
\cite{karol}, all of which are constructed with Dita's method. We
will now prove the following

\begin{proposition}  The matrix $S_{12}$ is not included (even up
to equivalence) in any of the known $12\times 12$ families listed in \cite{karol}.
\end{proposition}
\noindent{\it Proof.} By checking existing $I$-equivalences between rows (and columns)
it is elementary to show that $S_{12}$ (and its
transposed) is {\it not Dita-type}. To find possible index sets $I$ and $I$-equivalences between rows (resp. columns, in the transposed case) it is perhaps most convenient to note the position of 0's in $\log S_{16}$ and look for $2\times 6$, $3\times 4$, $4\times 3$ and $6\times 2$ blocks of 0's as in the last part of the proof concerning $S_8$. In this case such blocks do not exist at all which immediately implies that $S_{12}$ is not Dita-type. Therefore $S_{12}$ is not contained in any
of the Dita-type families in \cite{karol}.

We must also
show that it does not belong to the families stemming from
$F_{12}$, as listed in \cite{karol}: $F^{(9)}_{12A}$,
$F^{(9)}_{12B}$, $F^{(9)}_{12C}$, $F^{(9)}_{12D}$, $(F^{(9)}_{12B})^T$, $(F^{(9)}_{12C})^T$, $(F^{(9)}_{12D})^T$. The key observation is that in each of these families
some rows (and columns) are left without parameters. In
particular, in each of the above families either the 1st and 7th or the 1st, 5th and 9th rows
remain unchanged. Therefore, in any matrix contained in these
families there are either two rows which are (2)-6-equivalent,
or three rows which are pairwise (3)-4-equivalent. It is easy to check (by
a short computer program, rather than by hand) that there are no
such rows in $S_{12}$. This means that $S_{12}$ is indeed not
contained in any of the known $12\times 12$ orbits. $\hfill \square$

By applying the linear variation of phases method of \cite{karol}
we have been able to find the following 5-parameter affine family
stemming from $S_{12}$. (Again, we can claim that the members of
this family are not Dita-type in a neighbourhood of $S_{12}$. We
do not know whether this affine family is maximal or further
parameters can be introduced).

\beql{s12a}S^{(5)}_{12}(a,b,c,d,e)=S_{12}\circ EXP(\textbf{i}R^{(5)}_{12}(a,b,c,d,e)),  \ \ \ \mathrm{where}\eeq

\beql{r12}R^{(5)}_{12}(a,b,c,d,e)=
\left[\begin{array}{cccccccccccc}
 \bullet & \bullet & \bullet & \bullet & \bullet & \bullet & \bullet & \bullet & \bullet & \bullet & \bullet & \bullet \\
 \bullet & \bullet & \bullet & e & e & e & \bullet & \bullet & \bullet & e & e & e \\
 \bullet & \bullet & \bullet & d & d & d & \bullet & \bullet & \bullet & d & d & d \\
 \bullet & \bullet & \bullet & e & e & e & \bullet & \bullet & \bullet & e & e & e \\
 \bullet & \bullet & \bullet & d & d & d & \bullet & \bullet & \bullet & d & d & d \\
 \bullet & \bullet & \bullet & \bullet & \bullet & \bullet & c & c & c & c & c & c \\
 \bullet & a & b & \bullet & a & b & \bullet & a & b & \bullet & a & b \\
 \bullet & a & b & \bullet & a & b & \bullet & a & b & \bullet & a & b \\
 \bullet & a & b & \bullet & a & b & \bullet & a & b & \bullet & a & b \\
 \bullet & \bullet & \bullet & \bullet & \bullet & \bullet & \bullet & \bullet & \bullet & \bullet & \bullet & \bullet \\
 \bullet & \bullet & \bullet & \bullet & \bullet & \bullet & \bullet & \bullet & \bullet & \bullet & \bullet & \bullet \\
 \bullet & \bullet & \bullet & \bullet & \bullet & \bullet & c & c & c & c & c & c
\end{array}\right]
\eeq

\section{Conclusion}

In this paper we have used the connection between tiling of
Abelian groups and complex Hadamard matrices to
{\it recover the general construction of Dita} \cite{dita}, and also to
{\it obtain new parametric families of order 8, 12 and 16} which
complement the recent catalogue \cite{karol}. The construction of the new families was
based on a spectral-set analogue of a tiling method of Szab\'o \cite{szabo}. In principle, the method of \cite{szabo}
works in any finite Abelian group $G=\Z_{p_1q_1}\times \Z_{p_2q_2}\times \Z_{p_3q_3}$ and the corresponding spectral sets yield complex Hadamard matrices of size $p_1p_2p_3$ for any $p_1,p_2,p_3\ge 2$. It is not clear whether different choices of $q_1, q_2, q_3$ lead to non-equivalent matrices. In this paper we have only included the cases where $p_1p_2p_3\le 16$, and for which we could prove that the arising matrices are new and thus complement the catalogue \cite{karol}. The next smallest dimension in which the method works is $p_1p_2p_3=2\cdot 3\cdot 3=18$. Also, it would be interesting to see a {\it conceptual} proof that the Hadamard matrices constructed with this method are {\it never Dita-type} (for the matrices $S_8, S_{12}, S_{16}$ above we have proved this by a case-by-case analysis of the rows and columns).

The {\it correspondence between tiling and complex Hadamard matrices} is interesting in its own right and may well lead to new families of Hadamard matrices in the future. To achieve this, one would need any new tiling construction (different from that of \cite{nspec} and \cite{szabo} which have been used in this paper), and use the spectral set analogue of the construction to produce new Hadamard matrices.

Finally, let us emphasize that our results may find direct application in various problems of
quantum information theory, since previously unknown complex
Hadamard matrices allow to construct new teleportation and dense
coding schemes and to find previously unknown bases of maximally
entangled states.

\end{document}